\documentclass[a4paper]{article}
\usepackage{epsfig}
\usepackage{citesort}
\usepackage{amsmath}
\usepackage{amssymb}
\usepackage{color}
\usepackage{url}

\usepackage{algorithm}
\usepackage{algpseudocode}
\usepackage{amsthm}
\usepackage{fancybox}

\newlength{\figurewidth}
\newlength{\smallfigurewidth}

\newtheorem{theo}{Theorem}
\newtheorem{defi}{Definition}

\newtheorem*{prf}{Proof}
\def\qed{\hfill $\Box$} 
\newtheorem{lemm}{Lemma}

\newcommand{\tetr}[2]{{}^{#2}#1}
\newcommand{\slog}{\text{slog}_{\varphi}}
\newcommand{\slogt}{\text{slog}_{2}}

\def\fact{\varphi}
\def\minf{\varphi^*}
\def\rem{r}

\newcommand{\ch}[1]{\mathcal{C}(#1)}
\newcommand{\dec}[1]{F[#1]}
\newcommand{\decf}[1]{F_{\fact}[#1]}

\newcommand{\dect}[2]{T_{#1}[#2]}

\newcommand{\dectfn}{T_{\fact}[\nbar]}
\def\nbar{\bar{n}}
\newcommand{\lamb}[2]{\Lambda(#1,#2)}
\newcommand{\minl}[1]{\Lambda^*(#1)}

\def\ord{\mathcal{O}}

\newcommand{\lterm}{$\lambda$-term}
\newcommand{\lterms}{$\lambda$-terms}

\begin{document}

\title
{\large
\textbf{Compaction of Church Numerals for Higher-Order Compression}
}

\author{%
	Isamu Furuya$^{\ast}$ and Takuya Kida$^{\ast}$\\[0.5em]
{\small\begin{minipage}{\linewidth}\begin{center}
\begin{tabular}{ccc}
&$^{\ast}$Hokkaido University& \\
&Kita 14-jo, Nishi 9-chome, Kita-ku& \\
&Sapporo, 060-0814, Japan& \\
&\url{{furuya, kida}@ist.hokudai.ac.jp}&\\
\end{tabular}
\end{center}\end{minipage}}
}

\maketitle
\thispagestyle{empty}

\begin{abstract}
  In this study, we address the problem of compacting Church numerals.
  Church numerals appear as a representation of the repetitive part of data in higher-order compression.
  We propose a novel decomposition scheme for a natural number using tetration,
  which leads to a compact representation of \lterms~equivalent to the original Church numerals.
  For natural number $n$,
  we prove that the size of the lambda term obtained by the proposed 
	method is 	$\ord((\slogt{n})^{\log{n}/\log{\log{n}}})$.
  Moreover, we quantitatively confirmed experimentally 
	that the proposed method outperforms a binary expression of Church numerals 
	when $n$ is less than approximately $10000$.
\end{abstract}

\section{Introduction}
\label{sec:intro}

The goal of this study is to obtain a compact lambda term (\lterm) 
that is equivalent to the Church numeral for a given natural number.
Church numerals are unary representations of natural numbers using lambda notation.
Herein, for an integer $n$, the length of the Church numeral increases linearly with $n$.
Let $\ch{n}$ be the Church numeral for a natural number $n$;
then the lambda expression $\ch{n}$ is
$(\lambda fx. \overset{n}{\overbrace{(f(f(\cdots f(f}}\, x)\cdots ))$.
For a large $n$, decomposing and representing it as an equivalent expression may reduce the length of the \lterm\, for $n$.
For example, $n=500$ can be decomposed as $5 \times 10 \times 10$.
The \lterm\, corresponding to the expression is given as
$((\lambda pqfx.p~(q~(q~f))~x)$ $\ch{5}~\ch{10})$,
which is much shorter than $\ch{500}$.

Reducing the length of Church numerals is applied in data compression.
Kobayashi et al.~\cite{Kobayashi2012} proposed a compression method
called higher-order compression that uses \lterms\, as the data model.
Their method translates an input to a \lterm\, by inducing the input itself
and then encoding the obtained \lterm.
Since repeating patterns in the \lterm\, appear with Church numerals,
shortening them is important for data compression.
We refer to the task of shortening Church numerals as 
the \emph{compaction of Church numerals}.

In this study, 
we propose the \emph{Recursive tetrational partitioning} (RTP) method
to decompose a natural number using \emph{tetration}.
We also present an algorithm to perform RTP for a given natural number
and demonstrate that the obtained expression is translated into a compact \lterm.
Moreover, we prove that the length of the obtained \lterm\, is $\ord((\slogt n)^{\log n/\log\log n})$ in the worst case.
Although this is slightly worse than $\ord(\log{n})$,
it can be reduced to $\ord(\slog{n})$ in the best case, with $\fact < n$.


Yaguchi et al.~\cite{Yaguchi2014EfficientCompression} recently proposed an efficient algorithm for higher-order compression.
They utilized a simply typed $\lambda$-term for efficient modeling and encoding.
Differing from Kobayashi et al.'s approach wherein each context occurring more than once is extracted,
Yaguchi et al. extracted the most frequent context up to a certain size.
In \cite{Yaguchi2014EfficientCompression},
they state that the performance of their method is often better than the performance of grammar compression,
with regard to compression ratio.
We confirm that the proposed method tends to produce more compact \lterms\, 
for highly repetitive patterns
compared with the method proposed by Yaguchi et al..
Note that the proposed method can be easily incorporated into their algorithm.

\noindent
{\bf Contributions:~}
The primary contributions of this study are as follows.
\begin{enumerate}
\item
  For natural numbers,
  we propose a novel decomposition scheme called RTP,
  which leads to compact representation of \lterms\, 
		that is equivalent to the Church numerals of the numbers.
  Note that the proposed RTP differs from $n$-ary notation.
\item
  By incorporating RTP,
  we propose an algorithm to perform the compaction of $\ch{n}$.
  Moreover, we prove that the length of the \lterms~constructed by the algorithm
		is $\ord((\slogt n)^{\log n/\log\log n})$ in the worst case.
\item
  We implemented the proposed algorithm and conducted comparative experiments,
  and results show that the proposed method is superior to that of Yaguchi el al.,
  and is also superior to the binary expression of Church numerals when $n$ is less than approximately $10000$.
\end{enumerate}

The remainder of this paper is organized as follows.
In Sec.~\ref{sec:preliminary}, we review lambda notation, Church numerals, and tetration.
In Sec.~\ref{sec:proposed}, we define the proposed RTP method and present the translation algorithm using RTP.
We also prove the upper bound of the length of the \lterm\, produced by our algorithm.
In Sec.~\ref{sec:application}, we describe how our algorithm is applied to higher-order compression,
review related work, and present experimental results.
Conclusions are presented in Sec.~\ref{sec:conclusion}.

\section{Preliminary}
\label{sec:preliminary}

\subsection{Lambda terms}

\begin{defi}[Lambda terms and their sizes]\label{defi:lamb}
  Let $S=\{\lambda, ., (, )\}$ be the set of special symbols.
  Let $A$ be the set of characters in the input data,
  where we assume $A\cap S=\emptyset$.
  We call $A$ and $a \in A$ \emph{terminal alphabet} and \emph{terminal symbol}, respectively.
  Let $\Sigma$ be an alphabet such that $\Sigma \cap (A \cup S)=\emptyset$.
  We call $x \in \Sigma$ \emph{variable}.
  For $a\in A$ and $x\in \Sigma$, \emph{lambda terms} (\lterms) are 
	 defined recursively as follows:
  \begin{align*}
	({\rm i})~x \qquad ({\rm ii})~(\lambda x.M) \qquad ({\rm iii})~(M ~ N)
	  \qquad ({\rm iv})~a
  \end{align*}
  where $M$ and $N$ are \lterms.
  We denote the size of the \lterm\, $M$ as $\#M$, 
  and we define each of its lambda terms as follows:
  \begin{align*}
	\#x = \#a = 1,~~~~~
	\#(\lambda x.M) = \#M + 1,~~~~~
	  \#(M\:N) = \#M + \#N + 1.
  \end{align*}
\end{defi}

The definition of the size of a $\lambda$-term can be found in \cite{Kobayashi2012}.
We refer to (ii) and (iii) in Def.~\ref{defi:lamb} as \emph{$\lambda$-abstraction}
and \emph{functional application}, respectively.
Although condition (iv) is added for higher-order compression,
Def.~\ref{defi:lamb} is inherently the same as that of the lambda calculus.
Thus, hereafter, we use well-known lambda calculus omission rules,
such as the omission of parentheses and short notation of nested $\lambda$-abstractions.

We use the \emph{de Bruijn index}~\cite{DEBRUIJN1972381},
which is a simplified notation system for lambda expressions, in our algorithm.
which is one of the simplified notations of lambda expression.
In the de Bruijn index, a $\lambda$-abstraction is written using only $\lambda$,
and each variable is represented by an integer $i\ge 0$.
Variable $i$ is bound by $i$th outer $\lambda$-abstraction
if the abstraction exists; otherwise, the variable is free.
For example, the $\lambda$-terms $\lambda x.\lambda y.(x~(a~(x~y)))$ and
$\lambda w.\lambda z.(w~(a~(w~z)))$ have the same meaning.
Using the de Bruijn index, both are expressed as $\lambda.\lambda.(1~(a~(1~0)))$.

\subsection{Church numerals}


\begin{defi}[Church numerals]\label{defi:Cn}
  Let $n$ be a natural number.
  \emph{Church numerals} for $n$, denoted $\ch{n}$, are defined as follows:
  \vspace{-14pt}
  \begin{align*}
	  \ch{n} := (\lambda fx. \overbrace{(f \:(f \cdots (f}^{n} \: x) \cdots )).
  \end{align*}
  Next, let $n_1$ and $n_2$ be natural numbers.
  Then, each function of addition, multiplication, and exponentiation on $\ch{n_1}$ and $\ch{n_2}$
  are, respectively, defined as follows:
  \begin{eqnarray*}
	  \text{Addition} &{\rm Add}(n_1,n_2)=n_1+n_2
	&:=~(\lambda pqfx.p~f~(q~f~x))~\ch{n_1}~\ch{n_2},\\
	  \text{Multiplication} &{\rm Mul}(n_1,n_2)=n_1\cdot ~n_2
	&:=~(\lambda pqfx.p~(q~f)~x)~\ch{n_1}~\ch{n_2},\\
	  \text{Exponentiation} &{\rm Exp}(n_1,n_2)={n_1}^{n_2}~~~~
	&:=~(\lambda pqfx.q~p~f~x)~\ch{n_1}~\ch{n_2}.
  \end{eqnarray*}
\end{defi}

As can be seen, $\lambda$-abstractions in the above \lterms\ appear first and
are followed by Church numerals.
We refer to the former as \emph{function parts} and the latter as \emph{argument parts}.

\subsection{Tetration and super-logarithm}

\emph{Tetration} is known as the next hyper-operation after exponentiation.
In mathematics, tetration is defined as iterated exponentiation.
For any natural numbers $\fact$ and $i$,
the $i$th tetration of $\fact$ is denoted $\tetr{\fact}{i}$ and defined recursively as follows:
\begin{align*}
  \tetr{\fact}{i}~:=
	~1 ~~ ({\rm for}~ i=0), 
	~~\fact^{\tetr{\fact}{i-1}}~~({\rm for}~ i>0).
\end{align*}
For example, $\tetr{2}{1}=2$, $\tetr{2}{2}=4$, $\tetr{2}{3}=16$, and $\tetr{2}{4}=65536$.
Here, from Def.~\ref{defi:Cn}, 
the function of tetration on $\ch{n_1}$ and $\ch{n_2}$ is defined as follows:\\[-20pt]
\begin{eqnarray*}
	\text{Tetration} &{\rm Tet}(n_1,n_2)=\tetr{n_1}{n_2}~~~~
	&:=~(\lambda pfx.\overbrace{p~p\cdots p}^{n_2}~f~x)~\ch{n_1}.
\end{eqnarray*}


The following lemma is easily induced from this definition.
\begin{lemm}\label{lemm:tetr}
  For natural numbers $\fact$ and $i$,
  it holds that $\log_{\fact}{\tetr{\fact}{i}} = \tetr{\fact}{i-1}$.
\end{lemm}

The \emph{super-logarithm}, denoted as $\mathrm{slog}$, is one of the inverse operations of tetration.
For natural numbers $\fact$ and $i$,
it holds that $\slog \tetr{\fact}{i} = i$.
For positive numbers, the super-logarithm is essentially equivalent to the iterated logarithm,
i.e., it holds that $\log^{\ast} n = \lceil \mathrm{slog}_{e} n \rceil$ for any $n>0$.

\section{Proposed Method}
\label{sec:proposed}

\subsection{Proposed approach}\label{sec:our_approach}
We perform compaction of $\ch{n}$ in the following two steps.
\begin{description}
	\item[Step 1:] Decompose $n$ with natural number $\fact$ ($1<\fact<n$)
		into a numerical expression
	  that includes as much tetration and multiplication of $\fact$ as possible.
  \item[Step 2:] Translate the expression into a corresponding $\lambda$-term such that the translated \lterm\, includes at least a single functional part followed by $\ch{\fact}$.
\end{description}
For Step 1, we introduce the proposed RTP in Sec.~\ref{sec:rtp}, and
for Step 2, we present a translation algorithm in Sec.~\ref{sec:trans_algo}.

As stated in Sec.~\ref{sec:intro},
we may reduce $\#\ch{n}$ for a large number $n$.
For the running example,
it becomes $\#\ch{500}=1003$,
while 
$\#((\lambda pqfx.p~(q~f)~x)~\ch{5}~((\lambda pqfx.p~(q~f)~x)$ $\ch{10}~\ch{10}))=85$,
which corresponds to $5\times 10\times 10$.
Moreover, we can compress the $\lambda$-term by combining two function parts into a single part,
such as $((\lambda pqfx.p~(q~(q~f))~x)$ $\ch{5}~\ch{10})$ with a size of $51$.

For a natural number $n$,
there are many ways to achieve decomposition,
and the size of the \lterm\, changes depending on the employed approach.
Note that obtaining optimal decomposition is difficult;
therefore, in RTP, we employ a heuristic approach.

\subsection{RTP}
\label{sec:rtp}

We only consider numerical expressions such as
$F ::=  x~|~F+F~|~F\cdot F~|~F^F$
in BNF, where $x$ is an arbitrary natural number.
If the calculation of $F$ results in $n$, we denote it $\dec{n}$.
Here, the goal is
to obtain $\dec{n}$ such that the size of the corresponding $\lambda$-term becomes smaller.
Reducing the kind of natural numbers used in $\dec{n}$ is effective for compaction.
Therefore, we consider $\dec{n}$ using only $\fact\le n$, denoted $F_{\fact}$, 
such as $F_{\fact} ::= \fact ~|~F_{\fact}+F_{\fact}~|~F_{\fact}\cdot F_{\fact} ~|~{F_{\fact}}^{F_{\fact}}$.
Note that $F_{\fact}$ only derives a multiple of $\fact$.
Let $\rem = (n \bmod \fact)$ and $\nbar = n - \rem$.
Then, if $F_{\fact}$ derives $\nbar$, we denote it $\decf{\nbar}$.
Here, $\decf{\nbar}+\rem$ is a numerical expression that derives $n$ 
and includes at most two kinds of natural numbers $\fact$ and $\rem$.

To reduce the size of the $\lambda$-term,
reducing arithmetic operations appearing in $\decf{\nbar}$ is also effective
because the size of the $\lambda$-term increases with the number of arithmetic operations.
Next, we show how the proposed method achieves this reduction.

We partition $\nbar$ into an addition of tetrations with integer coefficients as follows:
\begin{align*}
	\nbar \rightarrow \tetr{\fact}{k}\cdot p_k+ \tetr{\fact}{k-1}\cdot p_{k-1}
	+ \cdots + \tetr{\fact}{1}\cdot p_1
\end{align*}
where $k$ is the maximum natural number such that $\tetr{\fact}{k}\le \nbar$ and
$p_i~(0\le i\le k)$ is the integer such that $0\le p_i < \tetr{\fact}{i+1}$.
The term including $\tetr{\fact}{0}$ does not appear in it
because $\nbar$ is divisible by $\fact$.
Then, we convert each term as follows:
\vspace{-10pt}
\begin{equation*}
	\tetr{\fact}{i}\cdot p_i \rightarrow \tetr{\fact}{i}\cdot (p_i - r_i)
	+ (\overbrace{\tetr{\fact}{i}+\cdots +\tetr{\fact}{i}}^{r_i})
\end{equation*}
where $r_i=(p_i \bmod \fact)$.
Moreover, let $\bar{p_i}=p_i - r_i$ and 
partition $\bar{p_i}$ recursively in the same way. 
As a result, we can convert $\nbar$ to $\decf{\nbar}$.
The above procedure is defined as follows.

\begin{defi}[RTP]\label{defi:tetr}
	Let $n$ and $\fact$ be natural numbers such that $0 < \fact \le n$, 
    and let $\rem = (n \bmod \fact)$ and $\nbar = n - \rem$.
	Then, we define RTP as follows, where
    the result derived by RTP is denoted $\dectfn$:
    \begin{align*}
		\dectfn = 
		\begin{cases}
			\nbar
			&({\rm if}~ \nbar \le \fact)\\[-10pt]
			\tetr{\fact}{k}\cdot(\dect{\fact}{p_k - (p_k \bmod \fact)})
			+ (\overbrace{\tetr{\fact}{k}
			+\cdots +\tetr{\fact}{k}}^{p_k \bmod \fact})\\[-10pt]
			~~~~~~~~~~~~~
			+ \cdots + \tetr{\fact}{1}\cdot(\dect{\fact}{p_1- (p_1 \bmod \fact)})
			+ (\overbrace{\tetr{\fact}{1}+\cdots +\tetr{\fact}{1}}^{p_{1} \bmod \fact})
			&({\rm otherwise})
		\end{cases}
    \end{align*}
	where $k$ is the maximum natural number such that $\tetr{\fact}{k}\le \nbar$ and
	$p_i~(0\le i\le k)$ is the integer such that $0\le p_i < \tetr{\fact}{i+1}$.
	Here, if $p_i = 0$ or 1, 
	we do not display the term or coefficient, respectively.
\end{defi}

For example, 
$\dectfn=\dect{2}{200}=\tetr{2}{3}\cdot (\tetr{2}{2}\cdot 2+\tetr{2}{2})+\tetr{2}{2}\cdot 2$
with $n=201$ and $\fact=2$.
In Def.~\ref{defi:tetr}, each coefficient $(p_i-(p_i \bmod \fact))$ is 
a multiple of $\fact$.
Thus, the remainder of each recursion step will always be $0$.
Therefore, no term includes $\tetr{\fact}{0}$ in $\dectfn$.
In addition, $\dect{\fact}{\nbar}$ is determined uniquely
relative to the given $n$ and $\fact$.

\newcommand{\FR}{\textit{FR}}
\subsection{Translation algorithm}
\label{sec:trans_algo}

When numeral expression $F$ is represented by functional representation,
we denote it $\FR(F)$.
In this representation,
${\rm Num}(x)$, ${\rm Add}(x,y)$, ${\rm Mul}(x,y)$, and ${\rm Exp}(x,y)$
correspond to $x$, $x+y$, $x\cdot y$, and $x^y$, respectively.
Then, the following holds:
\begin{align*}
  &\FR(\dectfn) :=
	{\rm Add}(\text{term}(k),
	({\rm Add}(\text{term}(k-1),(\cdots,({\rm Add}(\text{term}(2),\text{term}(1)))\cdots),
\end{align*}
where \\[-32pt]
\begin{align*}
  &\tetr{\fact}{i}:=
  {\rm Exp}(\overbrace{\fact,{\rm Exp}(\fact,\cdots{\rm Exp}(\fact,\fact}^{i})\cdots),
  ~\mathrm{rem}(p_i):=
	{\rm Add}(\overbrace{\tetr{\fact}{i},{\rm Add}(\tetr{\fact}{i},
	\cdots{\rm Add}(\tetr{\fact}{i},\tetr{\fact}{i}}^{p_i\bmod \fact})\cdots),\\
  &\text{and}~\mathrm{term}(i):=
	{\rm Add}({\rm Mul}(\tetr{\fact}{i},(\FR(\dect{\fact}{p_i-(p_i\bmod\fact)}))),
  \mathrm{rem}(p_i))).
\end{align*}

While we can represent any numerical expression
using naive substitution via Def.~\ref{defi:Cn},
the $\lambda$-term generated in this manner tends to be large.
Therefore, we designed an algorithm that enables generation of a compact $\lambda$-term
relative to $\dectfn$.
This algorithm is described in Algorithm \ref{alg:toLambda}, and
we denote the generated $\lambda$-term by $\lamb{\dectfn}{\rem}$
with $\dectfn + \rem$, the numerical expression of $n$.
Note that we denote lambda terms using the de Bruijn index in Algorithm \ref{alg:toLambda}.
With Algorithm \ref{alg:toLambda}, for example, 
$200=\dect{2}{200}+1$ 
is translated to 
$\lamb{\dect{2}{200}}{1}$, 
where
\begin{align*}
	\lamb{\dect{2}{200}}{1}=
	((\lambda.\lambda.\lambda.(2~2~2~(\lambda.\lambda.(\lambda.(5~5~(5~2)~(5~5~2~0))~0)~1)~
	(2~2~(2~1)~0)))~\ch{2})
\end{align*}
Thus, $\#\lamb{\dect{2}{200}}{1}=49$, which is much smaller than $\#\ch{201}=405$.

\setlength\textfloatsep{0pt}
\begin{algorithm}[tb]
\caption{Algorithm translating $\dectfn+\rem$ into a $\lambda$-term}
\label{alg:toLambda}
\begin{algorithmic}[1]
\Require $\dectfn, \rem$
\Ensure $\lamb{\dectfn}{\rem}$
	\State {\bf let} ${\rm Translate}(i,j,term)=$
	\State ~~~~{\bf match} $term$ {\bf with}
	\State ~~~~$|~{\rm Num}(term)\rightarrow 2 + i * 2 + j$
	\State ~~~~$|~{\rm Add}(lTerm, rTerm)\rightarrow$
	\State ~~~~~~~~{\bf let} ${\rm LeftSideOfAdd}(lTerm)=$
	\State ~~~~~~~~~~~~{\bf match} $lTerm$ {\bf with}
	\State ~~~~~~~~~~~~$|~({\rm Add}(x,y)$ {\bf or} ${\rm Mul}(x,y))$ {\bf as} 
		       	$term \rightarrow {\rm Translate}(i+1, j, term)$
	\State ~~~~~~~~~~~~$|~{\rm otherwise}$ {\bf as} 
		        	$term \rightarrow ({\rm Translate}(i+1, j, term),2+i)$
	\State ~~~~~~~~{\bf in}
	\State ~~~~~~~~{\bf let} ${\rm RightSideOfAdd}(rTerm)=$
	\State ~~~~~~~~~~~~{\bf match} $rTerm$ {\bf with}
	\State ~~~~~~~~~~~~$|~{\rm Add}(x,y)$ {\bf as} 
		            $term \rightarrow ({\rm Translate}(i+1, j, term),1+i)$
	\State ~~~~~~~~~~~~$|~{\rm Mul}(x,y)$ {\bf as} 
		            $term \rightarrow ({\rm Translate}(i+1, j, term),0)$
	\State ~~~~~~~~~~~~$|~{\rm otherwise}$ {\bf as} 
		        	$term \rightarrow ((({\rm Translate}(i+1, j, term),2+i),0)$
	\State ~~~~~~~~{\bf in}
	\State ~~~~~~~~$\lambda.({\rm LeftSideOfAdd}(lTerm), {\rm RightSideOfAdd}(rTerm))$
	\State ~~~~$|~{\rm Mul}(lTerm, rTerm)\rightarrow$
	\State ~~~~~~~~{\bf let} ${\rm RightSideOfMul}(rTerm)=$
	\State ~~~~~~~~~~~~~{\bf match} $rTerm$ {\bf with}
	\State ~~~~~~~~~~~~~$|~({\rm Add}(x,y)$ {\bf or} ${\rm Mul}(x,y))$ {\bf as} 
		   	       	$term \rightarrow \lambda.\lambda.({\rm Translate}(i, j+1, term),i)$
	\State ~~~~~~~~~~~~~$|~{\rm otherwise}$ {\bf as} 
		   	       $term \rightarrow {\rm Translate}(i, j, term)$
	\State ~~~~~~~~{\bf in}
	\State ~~~~~~~~$({\rm Translate}(i,j,lTerm),({\rm RightSideOfMul}(rTerm),1+i))$
	\State ~~~~$|~{\rm Exp}(lTerm, rTerm)\rightarrow$
		   	$({\rm Translate}(i,j,lTerm),{\rm Translate}(i,j,rTerm))$
	\State {\bf in}
	\State {\bf let} ${\rm TopLevel}(term)=$
	\State ~~~~{\bf let} ${\rm rem} =(\overbrace{1\:(1\cdots(1}^{r}\:0)\cdots )))$ {\bf in}
	\State ~~~~{\bf match} $term$ {\bf with}
	\State ~~~~$|~{\rm Add}(lTerm, rTerm)\rightarrow$
	\State ~~~~~~~~{\bf let} ${\rm LeftSideOfTopAdd}(lTerm)=$
	\State ~~~~~~~~~~~~~{\bf match} $lTerm$ {\bf with}
	\State ~~~~~~~~~~~~~$|~{\rm Add}(x,y)$ {\bf as} 
		   		   	$term \rightarrow {\rm Translate}(0, 0, term)$
	\State ~~~~~~~~~~~~~$|~{\rm Mul}(x,y)$ {\bf as} 
		   		   	$term \rightarrow {\rm Translate}(0, 0, term)$
	\State ~~~~~~~~~~~~~$|~{\rm otherwise}$ {\bf as} 
		      	  $term \rightarrow ({\rm Translate}(0, 0, term),1)$
	\State ~~~~~~~~{\bf in}
	\State ~~~~~~~~$({\rm LeftSideOfTopAdd}(lTerm), {\rm TopLevel}(rTerm))$
	\State ~~~~$|~{\rm Mul}(lTerm, rTerm)$ {\bf as} $term \rightarrow$
		      	$({\rm Translate}(0, 0, term), rem)$
	\State ~~~~$|~{\rm otherwise}(lTerm, rTerm)$ {\bf as} $term \rightarrow$
					$(({\rm Translate}(0, 0, term),1), rem)$
	\State {\bf in}
	\State $(\lambda.\lambda.\lambda.{\rm TopLevel}(\dectfn),\ch{\fact})$
\end{algorithmic}
\end{algorithm}

The $\lambda$-term generated by Algorithm \ref{alg:toLambda} 
is a functional application of a single $\lambda$-abstraction and $\ch{\fact}$.
Here, the $\lambda$-abstraction is considered
a folded function of the arithmetic operations included in $\dectfn$.

\begin{lemm}\label{lemm:size}
	Let $n$ and $\fact$ be natural numbers such that $\fact <n$,
	and let $\rem = (n \bmod \fact)$ and $\nbar = n-r$.
	We denote the number of additions, multiplications, and expressions
	in $\dectfn$ by $N_a$, $N_m$, and $N_e$, respectively.
    Then, we obtain:
	\begin{equation*}
		\#\lamb{\dectfn}{\rem} < 10N_a + 5N_m + 2N_e + 2\fact + 2\rem + 12.
	\end{equation*}
\end{lemm}
We omit the proof of Lemma~\ref{lemm:size} due to space constraints.

\begin{lemm}\label{lemm:ge8}
	Let $n$ and $\fact$ be natural numbers such that $\fact <n$,
	and let $\rem = n \bmod \fact$ and $\nbar = n-r$.
    Then,
    $\dectfn +\rem$ such that $\#\ch{n}>\#\lamb{\dectfn}{\rem}$ exists
	when $n > 8$.
\end{lemm}
\begin{prf}
  	We show that $\mint{\nbar}$ such that $\#\ch{n}>\#\lamb{\mint{\nbar}}{\rem}$ 
	exists for any $n>8$.
	Let $\fact_{half}$ be the integer such that $\fact_{half} = \lfloor n/2 \rfloor$, and
	let $\rem_{\fact_{half}}=n\bmod\fact_{half}$ and $\nbar = n - \rem_{\fact_{half}}$.
	Here, $\rem_{\fact_{half}}$ is at most $1$.
	If we decompose $n$ into the numerical expression such that
	$\dect{\fact_{half}}{\nbar} + \rem_{\fact_{half}} 
	= \fact_{half} + \fact_{half} + \rem_{\fact_{half}}$,
	then 
	$\#\lamb{\dect{\fact_{half}}{\nbar}}{\rem_{\fact_{half}}}
	= 2\fact_{half} + 16 + \rem_{\fact_{half}} \le n + 18$ holds.
	Therefore, because $\#\ch{n} = 2n+3$, Lemma~\ref{lemm:ge8} holds for $n > 15$.
    As shown in Table \ref{table:lemm:ge8:prf},
    it also holds for $8 < n \le 15$.
\qed
\end{prf}

\setlength\textfloatsep{0pt}
    \begin{table}[tb]
  		\begin{tabular}{|c|c|c|c|c|c|c|c|c|}
  		\hline
          $n$ & $\#\ch{n}$ &
          $\#\lamb{\mint{\nbar}}{\rem}$ &
          $n$ & $\#\ch{n}$ &
          $\#\lamb{\mint{\nbar}}{\rem}$ &
          $n$ & $\#\ch{n}$ &
          $\#\lamb{\mint{\nbar}}{\rem}$ \\
          \hline \hline
  		  9 & 21 & 20 &
          12 & 27 & 24 &
          15 & 33 & 28
          \\
  		  10 & 23 & 22 &
          13 & 29 & 26 &
          & &
          \\
          11 & 25 & 24 &
          14 & 31 & 28 &
          & &
          \\
  		\hline
      \end{tabular}
  		\caption{$\ch{n}=2n+3$ and 
		  $\#\lamb{\mint{\nbar}}{\rem}$ in $8 < n\le 15$}
  	\label{table:lemm:ge8:prf}
  	\end{table}

\subsection{Further compaction}
\label{sec:further_compaction}

Lemma \ref{lemm:ge8} implies that if $\minf > 8$,
we can convert $\lamb{\mint{\nbar}}{\rem}$ into a more compact $\lambda$-term
by applying RTP to $\minf$ and translating its result into a $\lambda$-term
using Algorithm \ref{alg:toLambda}.
This operation can be applied recursively 
while $\minf$, at each recursion step, is greater than $8$.
We denote the final $\lambda$-term obtained as a result as follows:
\begin{align*}
	(\lambda pfx.M_i)~\ch{\minf_{i+1}}
	&:=\lamb{\mint{\bar{\minf_i}}}{\rem_{\minf_i}}\\
	\minl{n}&:=(\lambda pfx.M_0)((\lambda pfx.M_1)\cdots
	((\lambda pfx.M_N)~\ch{\minf_{N+1}})\cdots),
\end{align*}
where $0\le i\le N,~\minf_0=n,~\rem_{\minf_i}=\minf_i\bmod\minf_{i+1}$, and
$\bar{\minf_i}=\minf_i-\rem_{\minf_i}$.


\begin{lemm}\label{lemm:fsize}
	Let $n$ and $\fact$ be natural numbers such that $\fact <n$,
	and let $\rem = (n \bmod \fact)$ and $\nbar = n-r$.
	Then, the size of the function part of $\lamb{\mint{\nbar}}{\rem}$ 
	is less than or equal to
	that of $\lamb{\dect{2}{\nbar}}{\rem}$.
\end{lemm}
\begin{prf}
	We demonstrate this by reduction to contradiction.
	We assume that the size of the function part of $\lamb{\mint{\nbar}}{\rem}$
	is larger than that of $\lamb{\dect{2}{\nbar}}{\rem}$.
	Relative to the size of the Church numerals appearing in the argument parts, 
	$\#\ch{2} \le \#\ch{\minf}$ holds.
	$\#\lamb{\dectfn}{\rem}$ follows
	the sum of the sizes of the function and argument parts.
	Therefore, 
	$\#\lamb{\mint{\nbar}}{\rem} > \#\lamb{\dect{2}{\nbar}}{\rem}$ follows the assumption.
	However, this contradicts $\lamb{\mint{\nbar}}{\rem}$ 
	being the minimum $\lambda$-term of $\lamb{\dectfn}{\rem}$.
	Thus, the assumption is incorrect and the proposition is proven.
\qed
\end{prf}

\begin{theo}\label{theo}
	$\ord(\#\minl{n})$ is $\ord((\slogt n)^{\log n/\log{\log n}})$
	with natural number $n$.
\end{theo}
\begin{prf}
	Let $\fact$ be a natural number such that $\fact <n$,
	and let $\rem = (n \bmod \fact)$ and $\nbar = n-r$.
	Then, $\ord(\#\lamb{\mint{\nbar}}{\rem})$ is the sum of
	$\ord$(the size of the function part) and
	$\ord$(the size of the argument part).
	First, we consider $\ord$(the size of the function part).
	By Lemma \ref{lemm:fsize}, it is bounded by 
	$\ord$(the size of the function part of $lamb{\dect{2}{\nbar}}{\rem}$).
	Here, $\dect{2}{\nbar}$ is as follows:
	\begin{align*}
	  \dect{2}{\nbar} = 
	      \tetr{2}{k}\dect{2}{\bar{p}_k}
	    + (\overbrace{\tetr{2}{k}}^{p_k \bmod 2})
	    + \tetr{2}{k-1}\dect{2}{\bar{p}_{k-1}}
	    + (\overbrace{\tetr{2}{k-1}}^{p_{k-1} \bmod 2})
	    + \cdots + \tetr{2}{1}\dect{2}{\bar{p}_{1}}
	    + (\overbrace{\tetr{2}{1}}^{p_1 \bmod 2})
	\end{align*}
	where $\bar{p_i}$ is $p_i - (p_i \bmod 2)$, which is a coefficient of $\tetr{\fact}{i}$
	for $1 \le i \le k$.
	Here, $p_i \bmod 2$ is at most 1.
	Then, $(\slogt{n})-1 < k \le \slogt{n}$
	holds relative to $k$.
	This indicates that $k$ is the maximum integer such that $k \le \slogt{n}$.
	By Lemma \ref{lemm:size}, in the function part, the size increments
	by addition, multiplication, and exponentiation are at most 10, 5, and 2, respectively.
	Therefore, the maximum size of each term $(\tetr{2}{i}\cdot\bar{p}_i+\tetr{2}{i})$ is
	$2i+5+(\text{the size of $\bar{p_i}$})+10+2i$ where $i\le k$ 
	and these terms appear at most $k$ times.
	Note that RTP partitions each $\bar{p_i}$ recursively.
	We denote the number of recursion times by $\rho$.
	The following holds relative to the upper bound of the size of the function part in
	$\lamb{\dect{2}{\nbar}}{\rem}$:
	\begin{align*}
		\ord((2k+5+(\text{the size of $\bar{p_k}$})+10+2k)\cdot k)
		= \ord(k^2+k+k(\text{the size of $\bar{p_k}$}))
		= \ord(k^{\rho}).
	\end{align*}
	Note that $i$ can be $k$ in each recursion step, therefore,
	by Lemma \ref{lemm:tetr}, the following holds:
	\begin{align*}
		(\tetr{2}{k})^{\rho} \le n
		\iff \rho \log{\tetr{2}{k}} \le \log{n},~~
		\rho \le \frac{\log{n}}{\log{\tetr{2}{k}}}
		< \frac{\log{n}}{\log{\tetr{2}{(\slogt{n})-1}}}
		= \frac{\log{n}}{\log{\log{n}}}.
	\end{align*}
	Thus,
	$\ord(k^{\rho}) = \ord((\slogt{n})^{\log{n}/\log{\log{n}}})$
	holds.

	Second, we consider $\ord$(the size of the argument part).
	If $\minf \le 8$, it is constant because $\ord(\#\ch{\minf}) = \ord(\#\ch{8})$.
	If $\minf > 8$,
	$\ch{\minf}$ is compacted recursively.
	By the above proof,
	the upper bound of the size of the $\lambda$-term result is
	$\ord((\slogt{\minf})^{\log{\minf}/\log{\log{\minf}}}+\#\ch{\minf_1})$
	where $\minf_1$ is $\fact_1$ 
	in minimum $\lamb{\dect{\fact_1}{\minf}}{(\minf - (\minf \bmod \fact_1))}$.
	Here, $\minf$ is clearly less than $n$.
	This is followed by
		$\ord((\slogt{\minf})^{\log{\minf}/\log{\log{\minf}}}) \le
		\ord((\slogt{n})^{\log{n}/\log{\log{n}}})$.
	If $\ch{\minf_1} > 8$, it is also converted recursively.
	However, a similar inequality holds in each recursion step and
	the final $\minf_{\rho}$ results in a constant such that $\minf_{\rho} \le 8$.
	Therefore, 
	$\ord(\#\minl{n})$ is $\ord((\slogt n)^{\log n/\log{\log n}})$.
\qed
\end{prf}

\section{Application to Higher-Order Compression}
\label{sec:application}

\subsection{Overview of higher-order compression}

In higher-order compression,
an input text is first represented as a $\lambda$-term 
where each terminal symbol is combined recursively by functional application.
For example, the text $ababc\$$ can be represented as $((a~b)~((a~b)~(c~\$)))$.
Here, there is an equivalency between two $\lambda$-terms $M$ and $N$
if they both result in the same calculated $\lambda$-term.
The main part of higher-order compression is to convert the input $\lambda$-term
to a more compact term with the remaining equivalency between them.


During higher-order compression processing,
a repetition pattern in the $\lambda$-term appears in the form of a Church numeral.
For example, the $\lambda$-term corresponding to the string "$abcabcabcabc\$$"
can be transformed into
  	$(\lambda fx.f(f(f(f~x))))(a~b~c)~\$ =\ch{4}~(a~b~c)~\$$.
If the repetition number becomes large, i.e., the Church numeral is large,
we can compress it by compaction of Church numerals.

\subsection{Related work}


Kobayashi et al.~\cite{Kobayashi2012} introduce a binary expression of Church numerals.
For $\ch{n}$, the size of the expression is $\Theta(\log{n})$.
Earlier, Mogensen~\cite{Mogensen2001} proposed a binary expression of Church numerals
and generalized to higher number-bases.
The method introduced by Kobayashi et al. is essentially the same as Mogensen's method.

In addition,
Yaguchi et al. also proposed a compaction method \cite{Yaguchi2014EfficientCompression}.
Their method treats the tetrational feature of \lterms~
and can compress them in super-logarithmic size order, similar to our method.
We refer to their method as the YKS algorithm in this paper.

\subsection{Experiment}

\begin{figure}[t]
  \begin{minipage}{0.5\hsize}
\begin{center}
  \includegraphics[width=76mm]{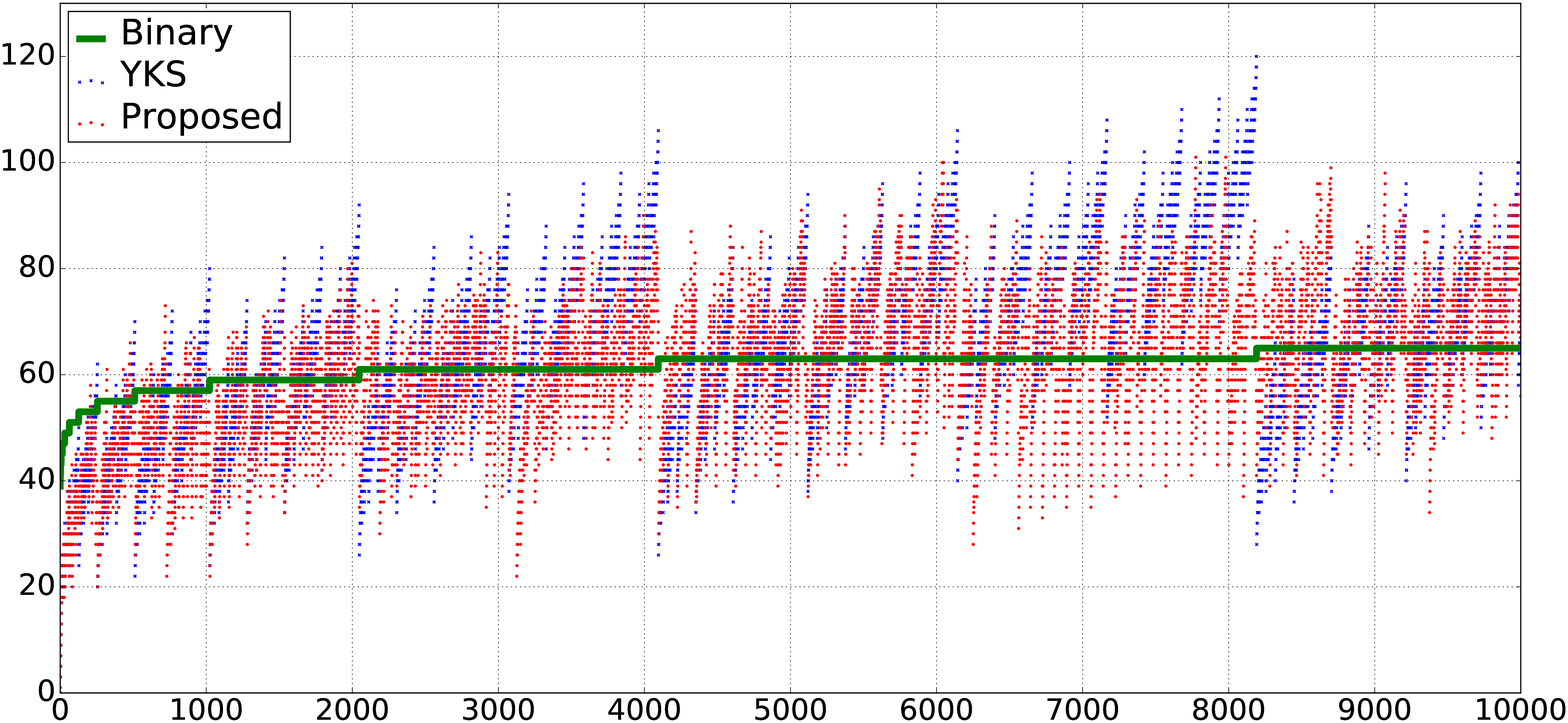}
\vspace{-14pt}
	\caption{Term size for integer $n$}
\label{fig1}
\end{center}
\end{minipage}
  \begin{minipage}{0.5\hsize}
\begin{center}
  \includegraphics[width=76mm]{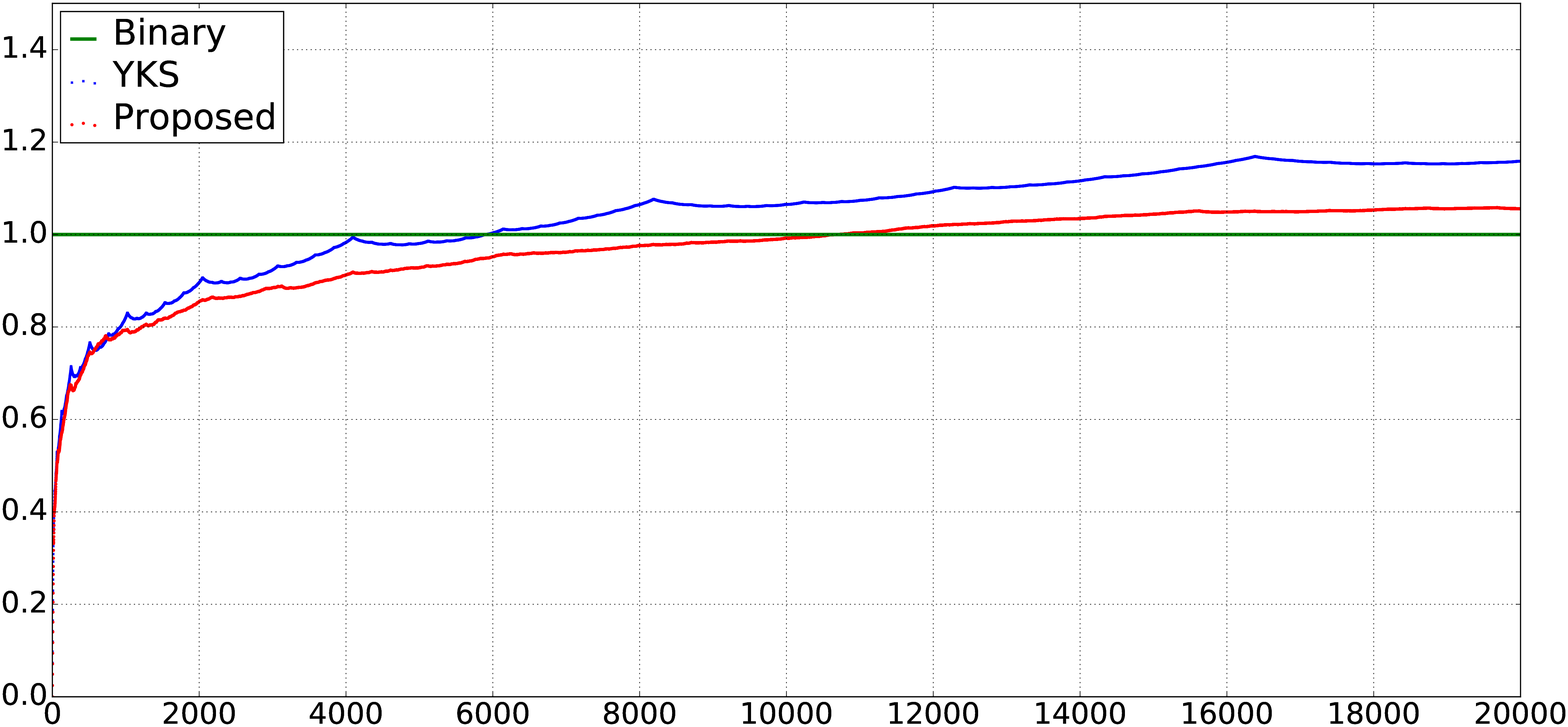}
\vspace{-14pt}
	\caption{Average ratio to binary expression}
\label{fig2}
\end{center}
\end{minipage}
\end{figure}

To compare the performance of the proposed method to that of the existing methods,
we conducted an experiment to evaluate the compression ratio.
Figures~\ref{fig1} and \ref{fig2} show the experimental results.
We used the artificial data $a^n\$$ as input.

\setlength\textfloatsep{0pt}


In Fig.~\ref{fig1}, the horizontal axis shows the repetition $n$
and the vertical axis shows the size of each $\lambda$-term.
Here, "Binary" denotes the size of the $\lambda$-term compacted 
by the method using the binary expression.
"YKS" denotes the YKS algorithm, and
"Proposed" denotes our proposed method.
The inequality (Proposed $\le$ Binary) holds in 5187 out of 10000 cases.
The average ratio (Proposed / Binary) is approximately 0.9962.
Similarly, the inequality (Proposed $\le$ YKS) holds in 5959 cases, and
the average ratio (Proposed / YKS) is approximately 0.9321.

In Fig.~\ref{fig2}, the horizontal axis shows the repetition $n$
and the vertical axis shows the ratio
((the average size of cumulative sum of $\lambda$-terms from $1$ to $n$) / Binary).
Fig.~\ref{fig2} shows how much the result of YKS and the proposed method will 
increase compared to the result of Binary 
if we assume that $n$ is given in uniform distribution.
As can be seen, the result of Proposed tends to be greater than 
  that of Binary when $n$ is greater than approximately $10000$.
We consider this to be consistent with the theoretical upper bound analysis 
	  result $\ord((\slogt n)^{\log n/\log\log n})$ stated in Theorem~\ref{theo}.

\section{Conclusion}
\label{sec:conclusion}

In this paper,
we have addressed the problem of compacting Church numerals,
which is useful for higher-order compression.
We have proposed RTP to decompose large numerals and presented a \lterm\, conversion algorithm using RTP.
We confirmed experimentally that
the \lterms\, produced by the algorithm have sizes
following its theoretical size, $\ord((\slogt n)^{\log n/\log\log n})$
in the worst case.
On the other hand, for $n=\tetr{2}{k}$, $\ch{n}$ is converted to a \lterm\, of $\ord(\slogt n)=\ord(k)$ size.

About bit encoding of \lterms, 
Tromp~\cite{tromp:DSP:2006:628} proposed a method for untyped \lterms.
In addition, very recently, Takeda et al.~\cite{TakedaCompactLambda-Terms}
proposed an efficient method to encode simply-typed \lterms.
Combining our method with these encodings is one of our future works.
Moreover, efficiently finding repeating regions in an input and counting the number of repetitions are remaining problems in higher-order compression.

\subparagraph*{Acknowledgments.}
The authors would like to thank Ayumi Shinohara and his colleagues for providing the source code for higher-order compression.
This work was supported by JSPS KAKENHI Grant Number JP15K00002
and JST CREST Grant Number JPMJCR1402, Japan.
In addition, the authors would like to thank Enago (www.enago.jp) for the English language review.


\begin{thebibliography}{1}
\providecommand{\url}[1]{#1}
\csname url@samestyle\endcsname
\providecommand{\newblock}{\relax}
\providecommand{\bibinfo}[2]{#2}
\providecommand{\BIBentrySTDinterwordspacing}{\spaceskip=0pt\relax}
\providecommand{\BIBentryALTinterwordstretchfactor}{4}
\providecommand{\BIBentryALTinterwordspacing}{\spaceskip=\fontdimen2\font plus
\BIBentryALTinterwordstretchfactor\fontdimen3\font minus
  \fontdimen4\font\relax}
\providecommand{\BIBforeignlanguage}[2]{{%
\expandafter\ifx\csname l@#1\endcsname\relax
\typeout{** WARNING: IEEEtran.bst: No hyphenation pattern has been}%
\typeout{** loaded for the language `#1'. Using the pattern for}%
\typeout{** the default language instead.}%
\else
\language=\csname l@#1\endcsname
\fi
#2}}
\providecommand{\BIBdecl}{\relax}
\BIBdecl

\bibitem{Kobayashi2012}
N.~Kobayashi, K.~Matsuda, A.~Shinohara, and K.~Yaguchi, ``Functional programs
  as compressed data,'' \emph{Higher-Order and Symbolic Computation}, vol.~25,
  no.~1, pp. 39--84, 2012.

\bibitem{Yaguchi2014EfficientCompression}
K.~Yaguchi, N.~Kobayashi, and A.~Shinohara, ``Efficient algorithm and coding
  for higher-order compression,'' in \emph{In proceedings of 2014 Data
  Compression Conference (DCC2014)}, March 2014, pp. 434--434.

\bibitem{DEBRUIJN1972381}
N.~de~Bruijn, ``Lambda calculus notation with nameless dummies, a tool for
  automatic formula manipulation, with application to the church-rosser
  theorem,'' \emph{Indagationes Mathematicae (Proceedings)}, vol.~75, no.~5,
  pp. 381 -- 392, 1972.

\bibitem{Mogensen2001}
T.~A. Mogensen, ``An investigation of compact and efficient number
  representations in the pure lambda calculus,'' in \emph{Revised Papers from
  the 4th International Andrei Ershov Memorial Conference on Perspectives of
  System Informatics: Akademgorodok, Novosibirsk, Russia}, ser. PSI '02.\hskip
  1em plus 0.5em minus 0.4em\relax London, UK: Springer-Verlag, 2001, pp.
  205--213.

\bibitem{tromp:DSP:2006:628}
J.~Tromp, ``Binary lambda calculus and combinatory logic,'' in \emph{Kolmogorov
  Complexity and Applications}, ser. Dagstuhl Seminar Proceedings, M.~Hutter,
  W.~Merkle, and P.~M. Vitanyi, Eds., no. 06051, 2006.

\bibitem{TakedaCompactLambda-Terms}
K.~Takeda, N.~Kobayashi, K.~Yaguchi, and A.~Shinohara, ``Compact bit encoding
  schemes for simply-typed lambda-terms,'' \emph{SIGPLAN Not.(Proceedings of
  the 21st ACM SIGPLAN International Conference on Functional Programming)},
  vol.~51, no.~9, pp. 146--157, Sep. 2016.

\end{thebibliography}

\end{document}